\documentclass[twocolumn,showpacs,preprintnumbers,amsmath,amssymb]{revtex4}
\usepackage{graphicx}
\usepackage{epsfig}
\usepackage{dcolumn}
\usepackage{bm}

\newcommand{\be}{\begin{enumerate}}
\newcommand{\ee}{\end{enumerate}}
\newcommand{\bi}{\begin{itemize}}
\newcommand{\ei}{\end{itemize}}
\newcommand{\mrecpipi}{m_{recoil}^{\pi^+\pi^-}}
\newcommand{\kskpipi}{K_s^0 K^{\pm}\pi^{\mp}\pi^0}
\newcommand{\kpipipi}{K^{\pm}\pi^{\mp}\pi^+\pi^-}
\newcommand{\opi}{\omega\pi^+\pi^-}
\newcommand{\oftwo}{\omega f_2(1270)}
\newcommand{\fipi}{\phi\pi^+\pi^-}
\newcommand{\fifo}{\phi f_0(980)}
\newcommand{\fikk}{\phi K^+ K^-}
\newcommand{\ok}{\omega K^+K^-}
\newcommand{\opp}{\omega p\bar p}  
\newcommand{\fipp}{\phi p\bar p}

\newcommand{\jpsi}{J/\psi}
\newcommand{\cbar}{\bar{c}}
\newcommand{\ppbar}{p \bar{p}}

\newcommand{\psipto}{\psi(2S)\rightarrow}
\newcommand{\jpsito}{J/\psi\rightarrow}
\newcommand{\psip}{\psi(2S)}
\newcommand{\rar}{\rightarrow}

\newcommand{\kkkk}{K^+K^-K^+K^-}
\newcommand{\pppp}{\pi^+\pi^-\pi^+\pi^-}
\newcommand{\ppp}{\pi^+\pi^-\pi^0}

\newcommand{\pipi}{\pi^+\pi^-}
\newcommand{\kk}{K^+K^-}

\newcommand{\ppb}{p\bar p}

\def\Journal#1&#2&#3(#4){#1{\bf #2}, #3 (#4)}

\def\NIMA{Nucl. Inst.  and Meths. {\bf A}}
\def\NPB{Nucl.  Phys.  {\bf B}}
\def\PLB{Phys.  Lett.  {\bf B}}
\def\PRL{Phys.  Rev.  Lett.  }
\def\PRD{Phys.  Rev.  {\bf D}}

\def\bec{\begin{center}}
\def\eec{\end{center}}

\begin{document}
\title{\Large \bf \boldmath $\psi(2S)$ two- and three-body hadronic decays }
\author{J.~Z.~Bai$^1$,      Y.~Ban$^{10}$,      J.~G.~Bian$^1$,
I.~Blum$^{18}$,     X.~Cai$^1$,       J.~F.~Chang$^1$,
H.~F.~Chen$^{17}$,  H.~S.~Chen$^1$,
J.~Chen$^4$,      Jie~Chen$^9$
% ...
\vspace{0.2cm}
\\(BES Collaboration)\\ 
\vspace{0.2cm}
$^1$ Institute of High Energy Physics, Beijing 100039, People's Republic of
     China\\
% ...
$^4$ Colorado State University, Fort Collins, Colorado 80523\\
% ...
$^9$ Nankai University, Tianjin 300071, People's Republic of China\\
$^{10}$ Peking University, Beijing 100871, People's Republic of China\\
% ...
$^{17}$ University of Science and Technology of China, Hefei 230026,
        People's Republic of China\\
$^{18}$ University of Texas at Dallas, Richardson, Texas 75083-0688\\
% ...
}

\noindent\vskip 0.2cm 
\begin{abstract}
We report  measurements of branching fractions for $\psip$ decays 
into $\opi$, $b_1\pi$, $\omega f_2(1270)$, $\ok$, $\opp$, $\fipi$, 
$\phi f_0(980)$ , $\fikk$, and an upper limit for $\fipp$ final states  based
on a data sample of $(4.02\pm0.22)\times10^6 ~ \psip$
events collected with the BESI detector at the Beijing
Electron-Positron Collider. The branching fractions for $b_1\pi$ and  
$\omega f_2(1270)$ update previous BES results, while those for other 
decay modes are first measurements. The ratios of $\psip$ and $\jpsi$ 
branching fractions are smaller than what is expected from the 12\% rule
by a factor of five for $\omega f_2(1270)$, 
by a factor of two for $\opi$, $\opp$, and $\fikk$,  
while for other studied channels the ratios are 
consistent with expectation within errors. \\ \\
\end{abstract}


\maketitle 

\section{Introduction}   \label{introd} 

In perturbative QCD, the charmonium states, $J/\psi$ and $\psi(2S)$,  
are considered to be non-relativistic bound states of charm and 
anticharm quarks, and their decays into light hadrons are expected to be 
dominated by the annihilation of the constituent $c$ and $\cbar$ quarks 
into three gluons. In this simple picture, 
the partial width for decays into any exclusive hadronic state $h$
 is proportional to the wave function at the origin squared, 
$|\psi(0)|^2$, which is well determined from dilepton decays. 
Since the strong coupling constant $\alpha_s$ does not change much 
between the $J/\psi$ and $\psip$ masses,
it is reasonable to expect that, for any 
exclusive hadronic state $h$, the $\jpsi$ and $\psip$ decay branching 
fractions will scale as \cite{qcd15} 
\begin{eqnarray*}
Q_h=
\frac{B(\psip\rar h)}{B(\jpsi\rar h)}
\simeq\frac{B(\psip\rar e^+e^-)}{B(\jpsi\rar e^+e^-)}\simeq 12\%,
\end{eqnarray*}
where the leptonic branching fractions are taken from the PDG 
tables \cite{PDG}. 
This relation is known 
as the ``$12\%$ rule''. Although the rule works reasonably well  for 
a number of specific decay modes, it fails severely in the case of the 
$\psip$ two-body decays to the vector-pseudoscalar ($VP$) meson final 
states, $\rho\pi$ and $K^*\bar K$ ~\cite{rhopi,ichep97}.  
This anomaly is commonly called the {\em $\rho\pi$ puzzle}.
In addition, the BES group has reported
violations of the $12\%$ rule for vector-tensor
($VT$) decay modes\cite{vt}.  Although a number
of theoretical explanations have been
proposed to explain this puzzle~\cite{qcd15,puzzle}, it seems that
most of them do not provide a satisfactory solution.

    In this paper, the measurements of the branching fractions of 
$\psip$ decays into 
$\opi$, $b_1\pi$, $\omega f_2(1270)$, $\ok$, $\opp$, $\fipi$,
$\phi f_0(980)$, $\fikk$, and $\fipp$ final states are presented.  The 
results are compared with the corresponding $\jpsi$ branching fractions to
test the $12\%$ rule for these two-body and three-body hadronic decays.

\section{The BES Detector}  \label{BESD} 
The BEijing Spectrometer, BES, is a conventional cylindrical  magnetic 
spectrometer that is coaxial with the colliding $e^+e^-$ 
beams of the Beijing Electron-Positron Collider, BEPC. BESI is described in 
detail in ref.~\cite{BES}. A four-layer central 
drift chamber (CDC) surrounding the beam pipe provides trigger information.
Outside the CDC, the forty-layer main drift chamber (MDC) provides 
tracking and energy-loss ($dE/dx$) information on charged tracks over  $85\%$
of the total solid angle.  The momentum resolution for charged tracks 
is $\sigma_p/p=0.017\sqrt{1+p^2}$ ($p$ in GeV/c), and the $dE/dx$ 
resolution for hadron tracks in these measurements is  about $9\%$.
An array of 48 scintillation counters surrounding the MDC provides 
measurements of the time-of-flight (TOF) of charged tracks with a
resolution of about 450 ps for hadrons. Outside  the TOF system is a 12 
radiation length thick lead-gas barrel shower counter (BSC) that 
operates in self-quenching streamer mode and detects
electrons  and photons over $80\%$ of the total solid angle.
The BSC energy resolution is $\sigma_E/E=0.22/\sqrt{E}$ ($E$ in GeV), 
and its spatial resolution for photons is $\sigma_\phi=4.5$ mrad 
and $\sigma_\theta=12$ mrad. A solenoidal magnet surrounds the BSC 
and provides a 0.4 Tesla magnetic field in the central tracking region of 
the detector. Outside the solenoidal coil, there are 
three double layers of proportional chambers interspersed with
the magnet flux return iron to identify muons of momentum greater 
than 0.5 GeV/c. 

\section{Event selection}  \label{selection} 
\subsection{Data sample and event topologies}  \label{method}
The data sample used for this analysis consists of 
$(4.02\pm0.22)\times10^6$ $\psip$ events
 collected with BES/BEPC at the center-of-mass energy 
$\sqrt s=M_{\psip}$.  The decay channels investigated are $\psip$ into 
$\opi$, $b_1\pi$, $\omega f_2(1270)$, $\ok$, $\opp$, $\fipi$,
$\phi f_0(980)$, $\fikk$, and $\fipp$ final states,
where  $b_1$ decays to $\omega\pi$, $\omega$ to $\ppp$, $\phi$ to $\kk$, and 
$f_2(1270)$ and $f_0(980)$ to $\pi^+\pi^-$.  They are all four-prong 
events or four-prong plus two photon events.

\subsection{Photon and charged particle identification} \label{pcid}
\begin{figure}[htbp]
\includegraphics[height=3cm,width=0.5\textwidth]{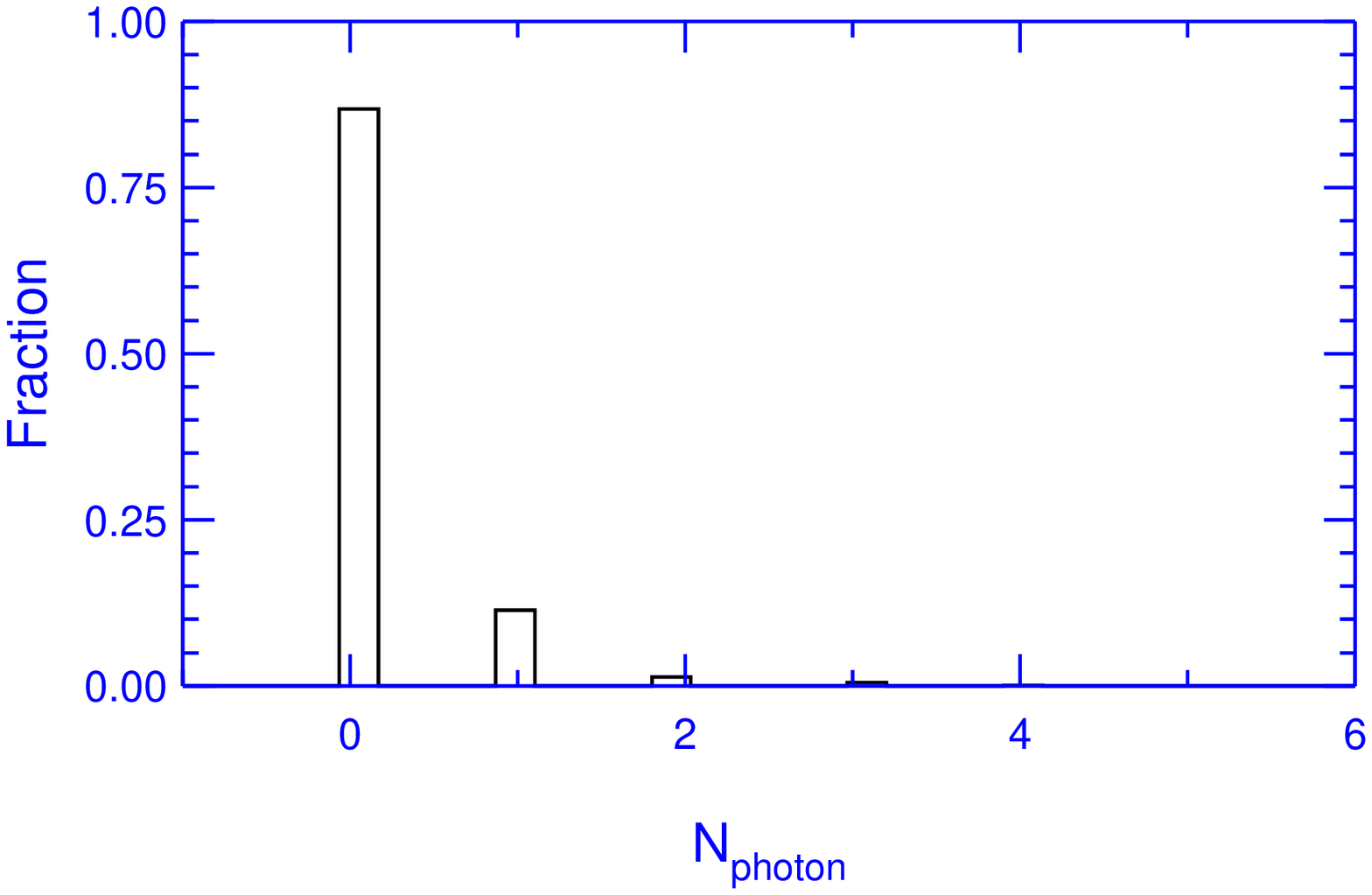}
\caption{\label{fig:photon}
 The distribution of the number of photon candidates found 
 in  kinematically selected $\psipto\pipi\ppbar$ events. }
\end{figure}

   A neutral cluster is considered to be a photon candidate if the following 
requirements are satisfied: it is located within the BSC fiducial region 
($|\cos\theta|<0.8$), the energy deposited in the BSC is greater than 50 
MeV, the first hit appears in the first 6 radiation lengths, 
the angle in the $x y$ plane (perpendicular to beam direction) between the 
cluster and the nearest charged track is greater than $16^\circ$, 
and the angle between the cluster development direction 
in the BSC and the photon emission direction from the beam interaction point
(IP) is less than $37^\circ$. With these criteria applied to 
the $\psipto\pipi\ppbar$ sample selected by four-constraint (4C) 
kinematic fitting, less than $20\%$ of 
events have photon candidates, which indicates an adequate fake-photon 
rejection (see Fig. \ref{fig:photon}).

  Each charged track is required  to be well fit by a three-dimensional
helix, to originate from the IP region, $V_{xy}=\sqrt{V_x^2+V_y^2}<2$ cm and
$|V_z|<20$ cm, and to have a polar angle $|\cos\theta|<0.8$. Here $V_x$, $V_y$,
and $V_z$ are the x, y, and z coordinates of the point of closest
approach to the beam axis.  The time of flight 
(TOF) and $dE/dx$ measurements for each charged track are used to 
calculate $\chi^2$ values for the hypotheses that a track is a pion,
kaon, or proton, for the purpose of particle identification.

\subsection{Monte Carlo simulations}  \label{MC}
   Phase space Monte Carlo (MC) event generators and the BES detector 
simulation package, SOBER~\cite{BES}, are used for simulating 
events for all channels analyzed.  
To determine detection efficiencies, MC generated events are subjected 
to the same reconstruction and event selection criteria as those applied to the
real data. For each channel, 30,000 MC events are generated.

\subsection{Event selection criteria}  \label{selec}
  For all analyzed decay channels, the candidate events are required 
to satisfy the following general selection criteria:
\renewcommand{\labelenumi}{\roman{enumi})}
\be
 \item The number of charged particles must be equal to four with net charge 
     zero.
 \item The number of photon candidates must be equal to or greater than two 
     for the decay channels containing $\pi^0$.  
 \item For each charged track in an event, the $\chi^2_{PID}(i)$ and 
its corresponding $Prob_{PID}(i)$ values are calculated based on the 
measurements of $dE/dx$ in the MDC and the time of flight in the TOF, with 
 definitions
$$\chi^{2}_{PID}(i)=\chi^{2}_{dE/dx}(i)+\chi^{2}_{TOF}(i)$$
$$Prob_{PID}(i)=Prob(\chi^{2}_{PID}(i),ndf_{PID}),$$
where $ndf_{PID}=2$ is the number of degrees of freedom in the $\chi^{2}_{PID}(i)$
determination and $Prob_{PID}(i)$ signifies the probability of this track 
having a particle $i$ assignment. For final states containing $\ppbar$, we 
require at least one of the charged tracks satisfy 
$Prob_{PID}(p/\overline{p})>0.01>Prob_{PID}(\pi/K)$, while for other 
channels analyzed, the
probability of a charged track for a candidate particle assignment 
is required to be greater than 0.01.
 \item A 4C (4 prong events) or 5C (4 
prong plus two photon events) kinematic fit is performed for each event. 
To be selected for any candidate final state, the event probability
given by the fit must be greater than 0.01.
 \item The combined $\chi^2$, $\chi_{com}^{2}$, is defined as the sum
of the $\chi^2$ values of the kinematic fit and those from
each of the four particle identification assignments: 
$$\chi_{com}^{2}=\sum_{i}\chi^{2}_{PID}(i)+\chi^{2}_{kine},$$
which corresponds to the combined probability:
$$Prob_{com}=Prob(\chi_{com}^2,ndf_{com}),$$
where $ndf_{com}$ is the corresponding total number of degrees of the
freedom in the $\chi_{com}^{2}$ determination.
The final state with the largest $Prob_{com}$ is taken as the  
candidate assignment for each event.
 \item A cut on $R_{Ep}$ is imposed to reject possible contamination 
from $\psipto\pipi\jpsi$ and $\eta\jpsi$, with $\jpsi\rar e^+e^-$, where
$$R_{Ep}=(\frac{E_{sc}^+}{p_+}-1)^2+(\frac{E_{sc}^-}{p_-}-1)^2,$$
and  $p_+$ ($p_-$)is the momentum of positive (negative) charged track measured with the 
MDC, and $E_{sc}^+$ ($E_{sc}^-$) is the energy deposited in the BSC by
the positive (negative) charged track. 
 \item Hit information from the muon chambers is used to reject possible 
muon tracks to reduce contamination from 
$\psipto\pipi\jpsi$ and $\eta\jpsi$, where $\jpsi\rar\mu^+\mu^-$.   
\ee
\subsubsection{$\psipto\opp$}
    The combined probability for the assignment of $\psipto\ppp\ppb$  
is required to be larger than those of $\psipto\ppp\pipi$ and 
$\psipto\ppp\kk$.
We impose a 
cut of $|\mrecpipi-m_{J/\psi}|>0.05$ GeV to reject backgrounds from 
$\psipto\pipi\jpsi$, where $\mrecpipi$ is the mass recoiling against the 
assigned $\pipi$ pair.  A requirement of
$m_{\ppbar}<m_{\psip}-m_{\omega}=2.9$ GeV
is applied to reject the backgrounds from 
$\psipto\eta\jpsi\rar\ppp\ppb$ and $\ppp\mu^+\mu^-$. Possible
background could come from
the decay of $\psipto\pi^0\pi^0\jpsi$, $\jpsi\rar p\bar{p}\pipi$, where one 
of the $\pi^{0}$s is missed in the BES detector. 
However, MC simulation shows that after our selection criteria, the $\ppp$ system 
from this process has a negligible contribution
in the $\omega$ mass region.  Also, due to the tiny branching fraction, 
the contamination from the decay of $\psipto\pi^0\jpsi$, $\jpsi\rar\pipi\ppb$ 
is negligible. 
     
\begin{figure}[htbp]
\centerline{\epsfig{figure=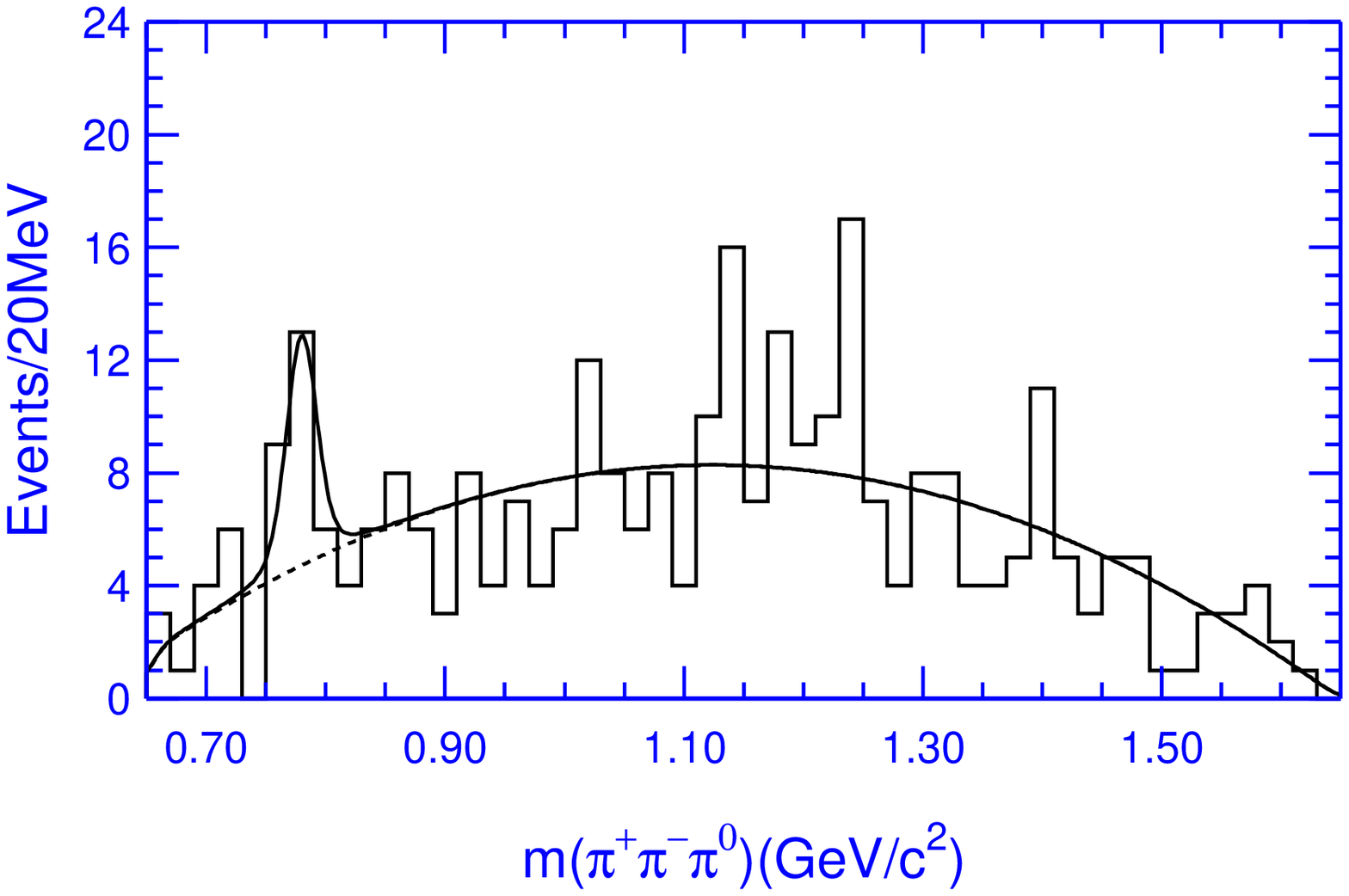,height=4.5cm,width=0.5\textwidth}}
\caption{\label{fig:wppfit}
  The $\ppp$ invariant mass distribution for
  candidate $\psipto\opp$ events. }
\end{figure}

 The $\ppp$ invariant mass distribution for the events that survive all 
selection requirements is shown in Fig. \ref{fig:wppfit}, where a 
clean $\omega$ signal can be seen. A Breit-Wigner resonance convoluted 
with Gaussian mass resolution function plus a polynomial background is fitted
to the data using an unbinned maximum likelihood method. In the fit, the 
mass resolution is fixed to its MC-determined value, and the width of 
the $\omega$ is fixed to its PDG value.  The fit gives 
$14.9\pm5.8$ signal events with statistical significance $2.6\sigma$.
In terms of  MC-determined efficiency of $5.4\%$, we 
determine the branching fraction
$$B(\psipto\opp)=(0.8\pm0.3\pm0.1)\times 10^{-4},$$
where the first error is statistical and the second error  
systematic.  Determination of the systematic errors is 
described in section IV.
 
\subsubsection{$\psipto\ok$}
        For this channel, the final state ($\ppp\kk$) is similar to 
that of the previous channel ($\ppp\ppb$) except $\ppb$ is replaced by $\kk$. 
Therefore, similar selection criteria are 
imposed, but the combined probability for the assignment of $\psipto\ppp\kk$
must be larger than those of $\psipto\ppp\pipi$ and
$\psipto\pi^0\kk\kk$.
A cut of $|\mrecpipi-m_{J/\psi}|>0.05$ GeV is used to reject backgrounds from
$\psipto\pipi\jpsi$. We require $m_{\kk}<m_{\psip}-m_{\omega}=2.9$ GeV
to reject backgrounds from $\psipto\eta\jpsi\rar\ppp\kk$.
The contamination from the decay of $\psipto\pi^0\jpsi,\jpsi\rar\pipi\kk$
is negligible  due to its tiny branching fraction.  Although our selection 
criteria can not completely eliminate the contamination from 
$\psipto\kskpipi$, $K_s^0\rar\pipi$ decay, the invariant mass
distribution of 
$m_{\ppp}$ from this background is smooth, and therefore it will not affect the 
determination of the signal events.

\begin{figure}[htbp]
\centerline{\epsfig{figure=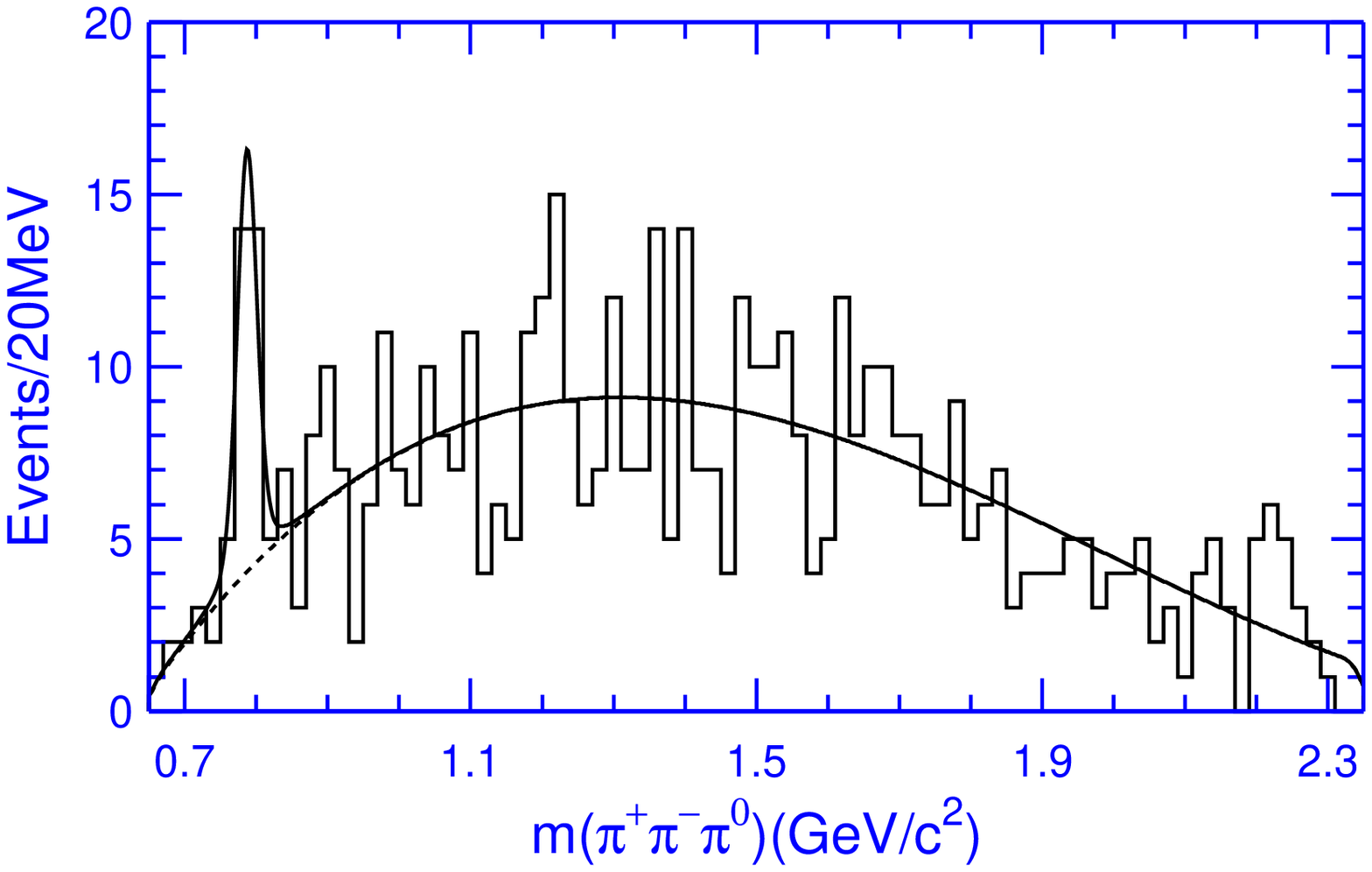,height=4.5cm,width=0.5\textwidth}}
\caption{\label{fig:wkkfit} The $\ppp$ invariant mass distribution for
candidate $\psipto\ok$ events. }
\end{figure}
Figure \ref{fig:wkkfit} shows the $\ppp$ invariant mass distribution for
$\ok$ candidates. The polynomial backgrounds include the
contamination from $\psipto\kskpipi$, $K_s^0\rar\pipi$.  A fit gives
$23.0\pm5.2$ signal events with a statistical significance of 6.3 
$\sigma$. The detection efficiency for this
decay mode is $4.4\%$, and we determine the branching fraction
$$B(\psipto\ok)=(1.5\pm0.3\pm0.2)\times 10^{-4}.$$

\subsubsection{$\psipto\opi$}
    The candidate events for this decay 
mode have the final state $\ppp\pipi$.  To be selected, the combined probability for the assignment of $\psipto\ppp\pipi$
must be larger than that of $\psipto\ppp K^+K^-$.
A cut of $|\mrecpipi-m_{J/\psi}|>0.05$ GeV  rejects the backgrounds from
$\psipto\pipi\jpsi$, $\jpsito\ppp$. We require 
$m_{\pipi}<m_{\psip}-m_{\omega}=2.9$ GeV
to reject the backgrounds from 
$\psipto\eta\jpsi\rar\ppp\pipi$ and $\ppp\mu^+\mu^-$, where  
$m_{\pipi}$ is the invariant mass of the $\pipi$ against the $\omega$ 
determined by the kinematic fit.
The contamination from the decay of $\psipto\pi^0\jpsi$, $\jpsi\rar\pppp$
is negligible  due to its tiny branching fraction.

\begin{figure}[htbp]
\centerline{\epsfig{figure=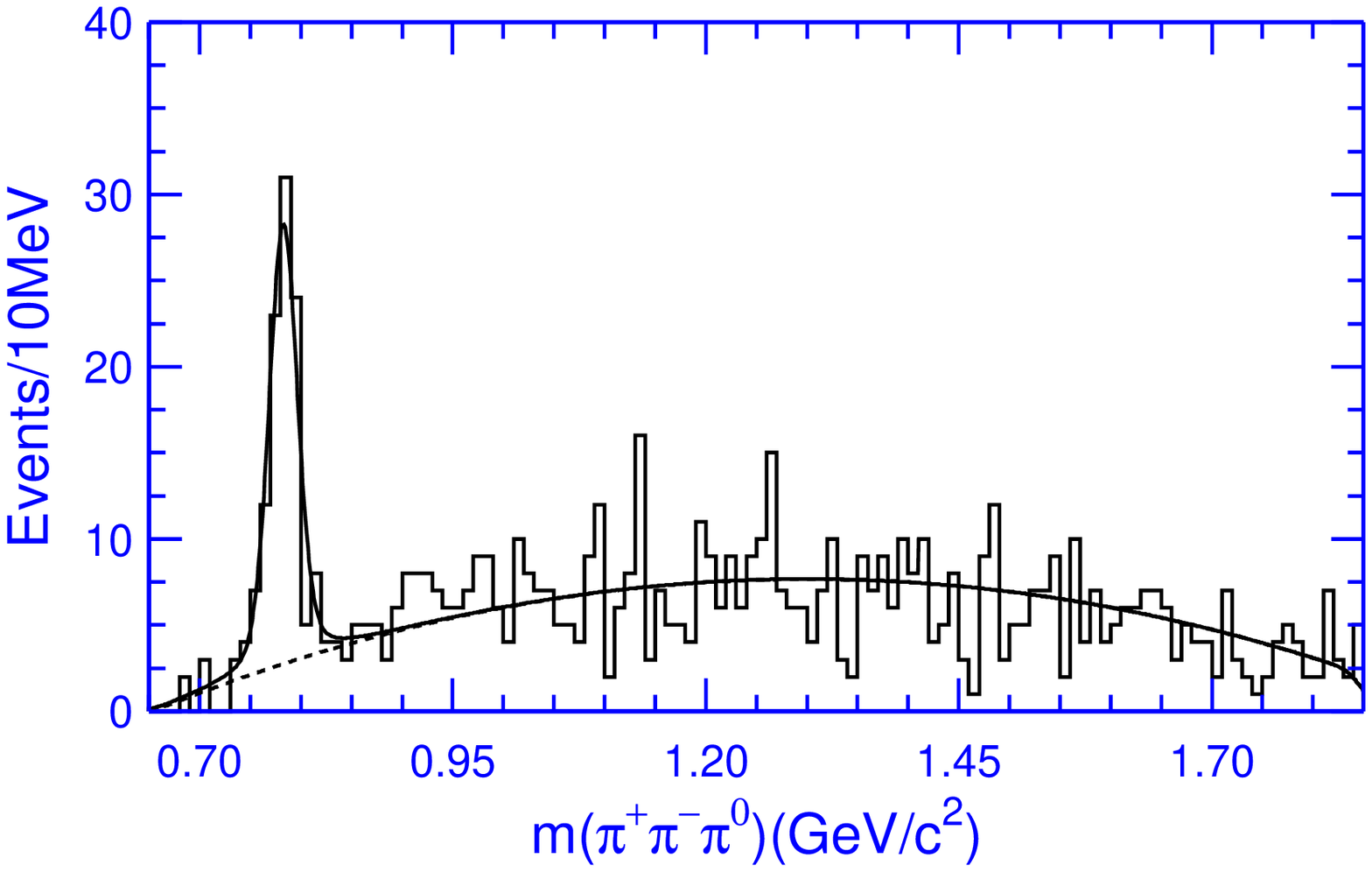,height=4.5cm,width=0.5\textwidth}}
\caption{\label{fig:wpipifit} 
  The $\ppp$ invariant mass distribution for
  candidate $\psipto\opi$ events. }
\end{figure}

Figure \ref{fig:wpipifit} shows the the $\ppp$ invariant mass distribution for
$\opi$ candidates, where the polynomial backgrounds contain the
contamination from $\psipto\kskpipi$, $K_s^0\rar\pipi$.  A fit gives
$100\pm12$ signal events. The detection efficiency for this
decay mode is $5.8\%$, and we determine the branching fraction
$$B(\psipto\opi)=(4.8\pm0.6\pm0.7)\times 10^{-4}.$$

\subsubsection{$\psipto b_1\pi$}
\begin{figure}[htbp]
\centerline{\epsfig{figure=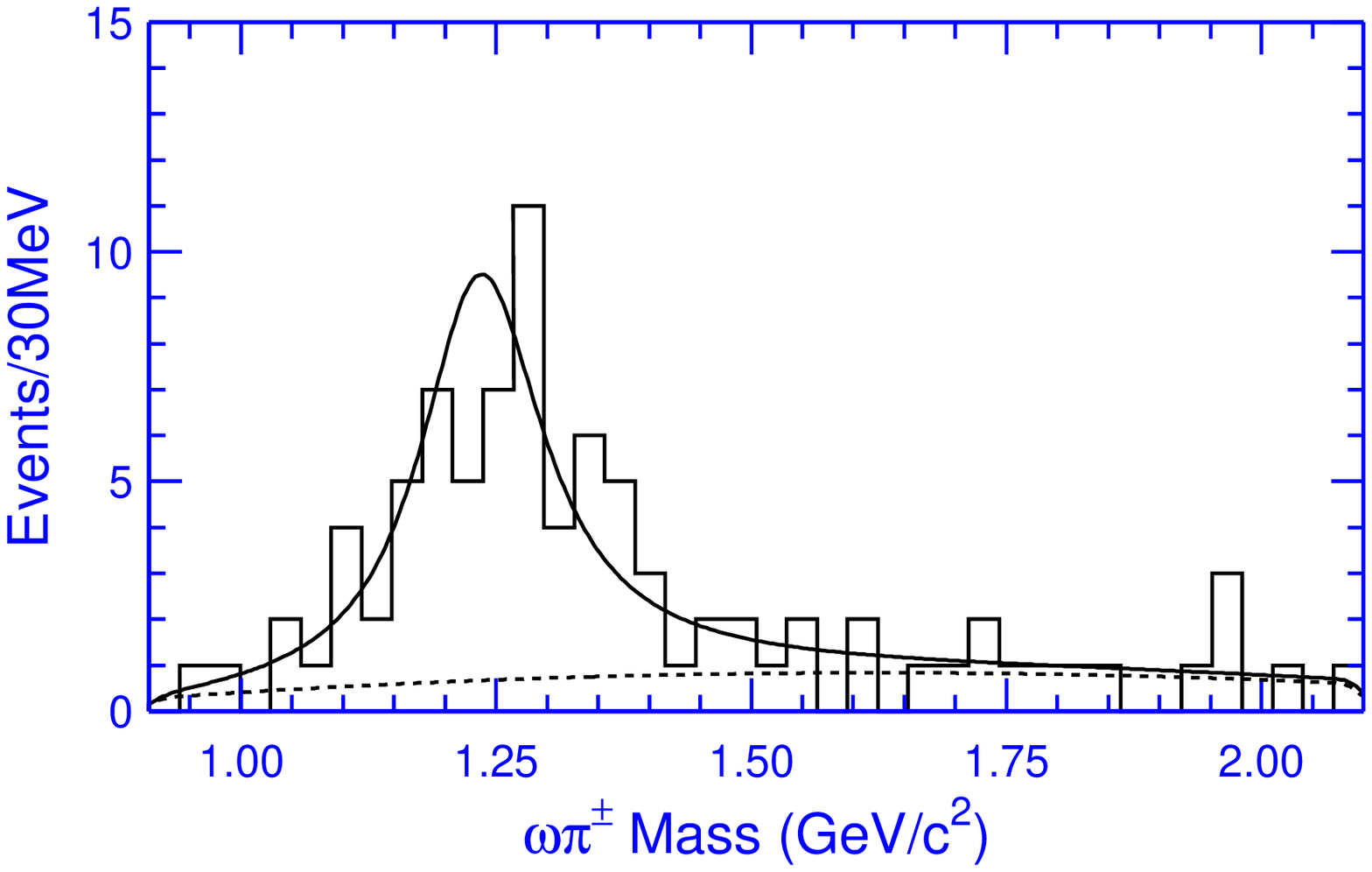,height=4cm,width=0.5\textwidth}}
\caption{\label{fig:b1pifit} 
  The $\omega\pi$ invariant mass distribution for
  candidate $\psipto b_1\pi$ events. }
\end{figure}

\begin{figure*}
\includegraphics[height=8cm,width=13cm]{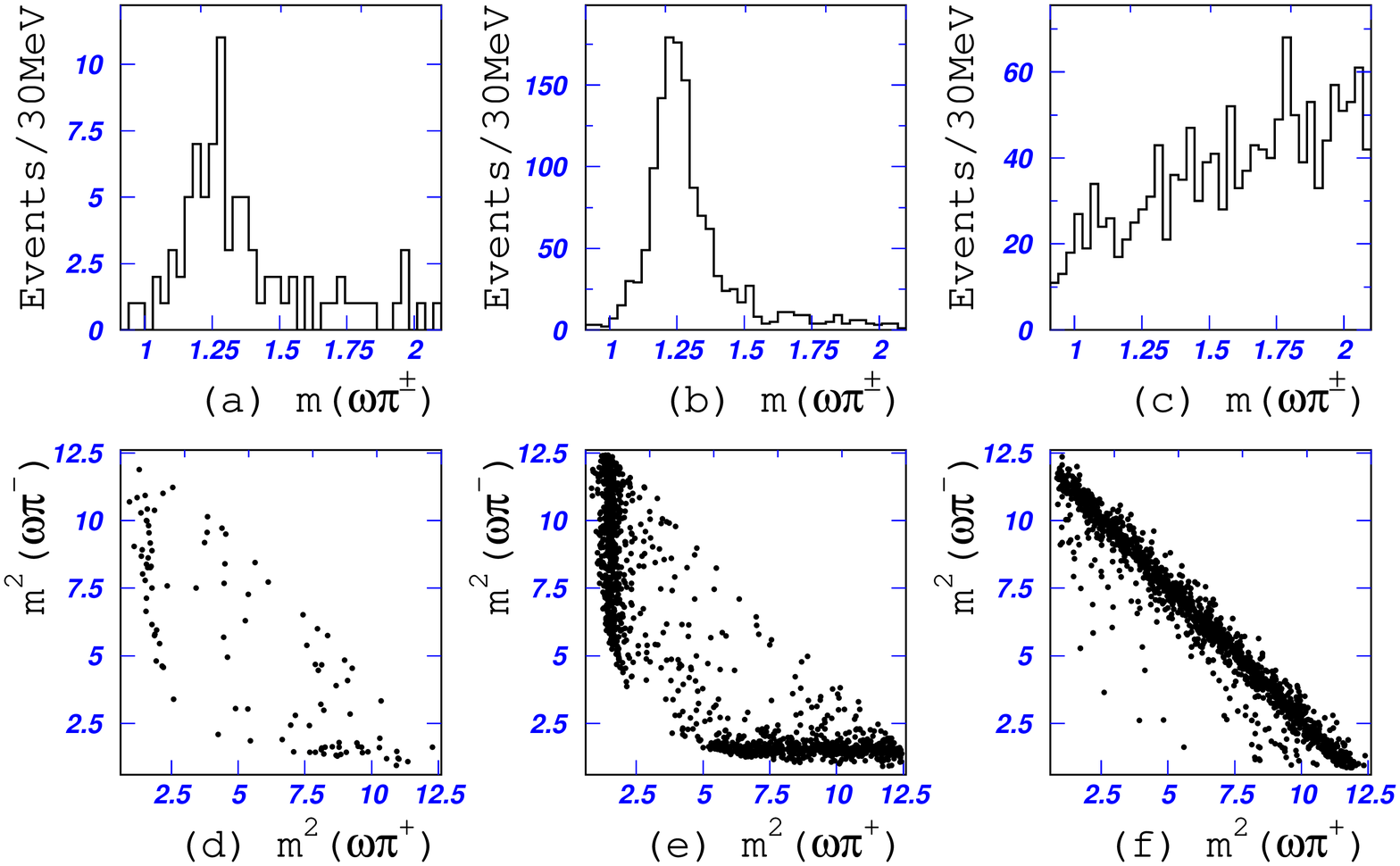}
\caption{\label{fig:b1pidalitz}
	  The invariant mass $m(\omega\pi^{\pm})$ and 
	   Dalitz plot (a,d) for $\psip\rar \opi$ (data); 
	  (b,e) for $\psipto b_1\pi$ (MC); and (c,f) for 
	  $\psipto \omega f_2(1270)$ (MC) events, respectively. }
\end{figure*}

    The dominant decay mode of the $b_1$ is $b_1\rar\omega\pi$, 
and we assume its branching fraction is $100\%$. Therefore, the final 
state for this mode is the same as for $\psipto\opi$.  We use the same criteria 
as those for $\psipto\opi$ to select candidate events, but an additional cut
$|m_{\ppp}-m_{\omega}|<0.03$ GeV is applied to select events
containing the $\omega$ particle. The Dalitz plot  is shown in 
Fig.~\ref{fig:b1pidalitz}.  The dense clusters in the top-left and in
the bottom-right of the scatter plots (d) and (e) indicate a clear $b_1$ signal.     
Figure \ref{fig:b1pifit} shows the $\omega\pi$ invariant mass 
distribution for $b_1\pi$ candidates. In the fit, the mass and width of 
the $b_1$ are fixed to the PDG values. A fit gives $61\pm11$ signal 
events with statistical significance $6.6 \sigma$. The detection efficiency for this
decay mode is $5.2\%$, and we determine the branching fraction
$$B(\psipto  b_1\pi)=(3.2\pm0.6\pm0.5)\times 10^{-4}.$$

\subsubsection{$\psipto \oftwo$}
\begin{figure}[htbp]
\centerline{\epsfig{figure=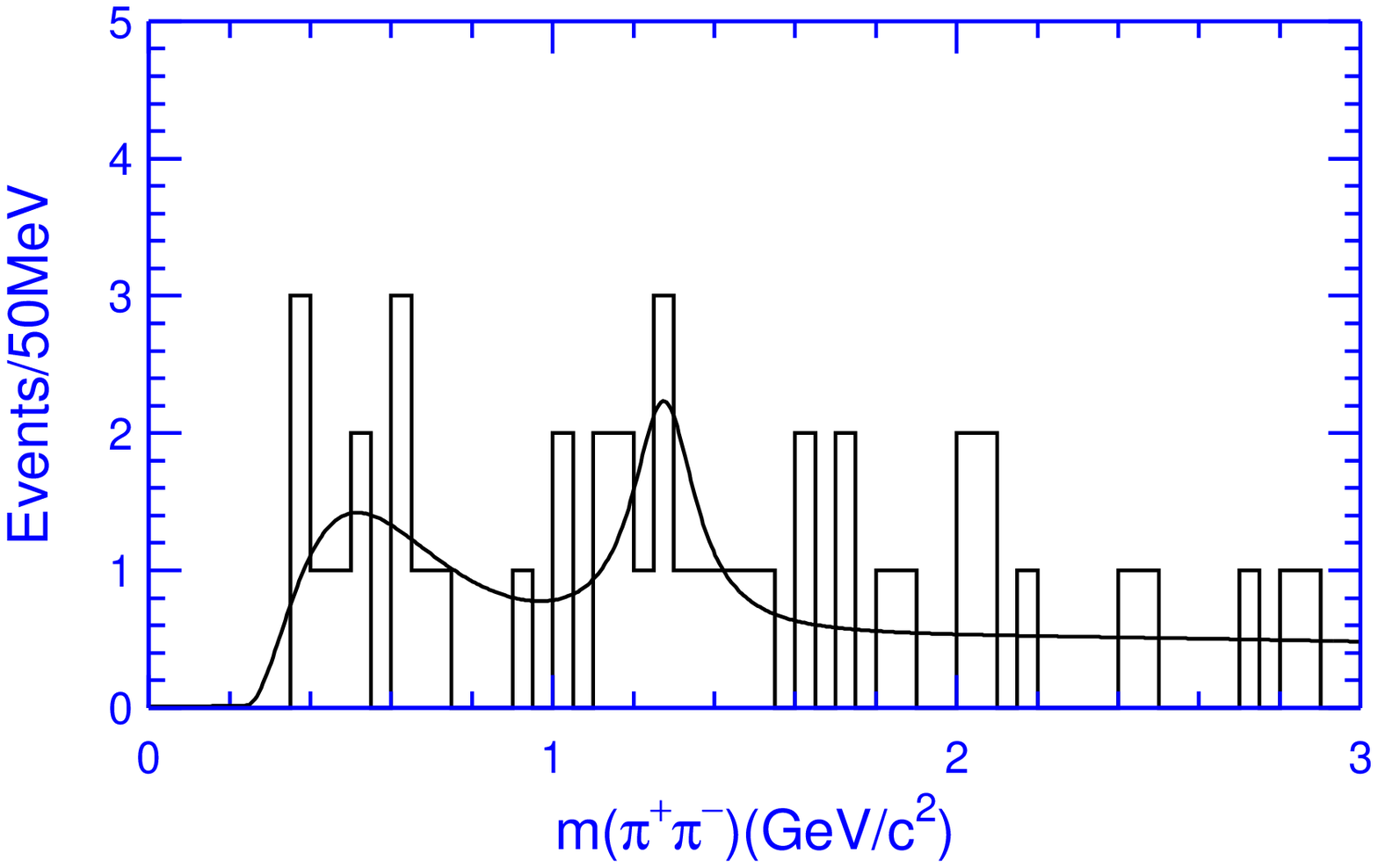,height=4cm,width=0.5\textwidth}}
\caption{\label{fig:wf2fit}
	The  $\pipi$ invariant mass distribution for candidate 
	$\psipto\oftwo$ events. }
\end{figure}

    The final state for this decay mode is also the same as for $\psipto\opi$.  
We use the same criteria as those for $\psipto\opi$, but impose an additional 
cut $|m_{\ppp}-m_{\omega}|<0.03$ GeV to select events containing an
$\omega$ particle.  A requirement of  $|m_{\omega\pi}-m_{b_1}|>0.2$ GeV 
is applied to remove contamination from the $b_1\pi$ channel. 
Figure \ref{fig:wf2fit} shows the $\pipi$ invariant mass
distribution for $\psipto\oftwo$ candidates; it shows a visible bump in 
the $f_2(1270)$ mass region, in addition to the broad distribution in
the lower mass 
region, which is presumably attributed to $f_0(400-1200)$ 
~\cite{sigma} production.
A fit gives $10.2\pm4.9$ signal events with the mass and width of
the $f_2(1270)$ fixed to its PDG values,
the statistical significance is about $2.1\sigma$.
The detection efficiency for this  
decay mode is $4.8\%$, and we determine the branching fraction
$$B(\psipto\oftwo)=(1.1\pm0.5\pm0.2)\times 10^{-4},$$
or an upper limit of $1.5\times 10^{-4}$ (90\% C.L.).

\subsubsection{$\psipto \fipi$}  
\begin{figure}[htbp]
\centerline{\epsfig{figure=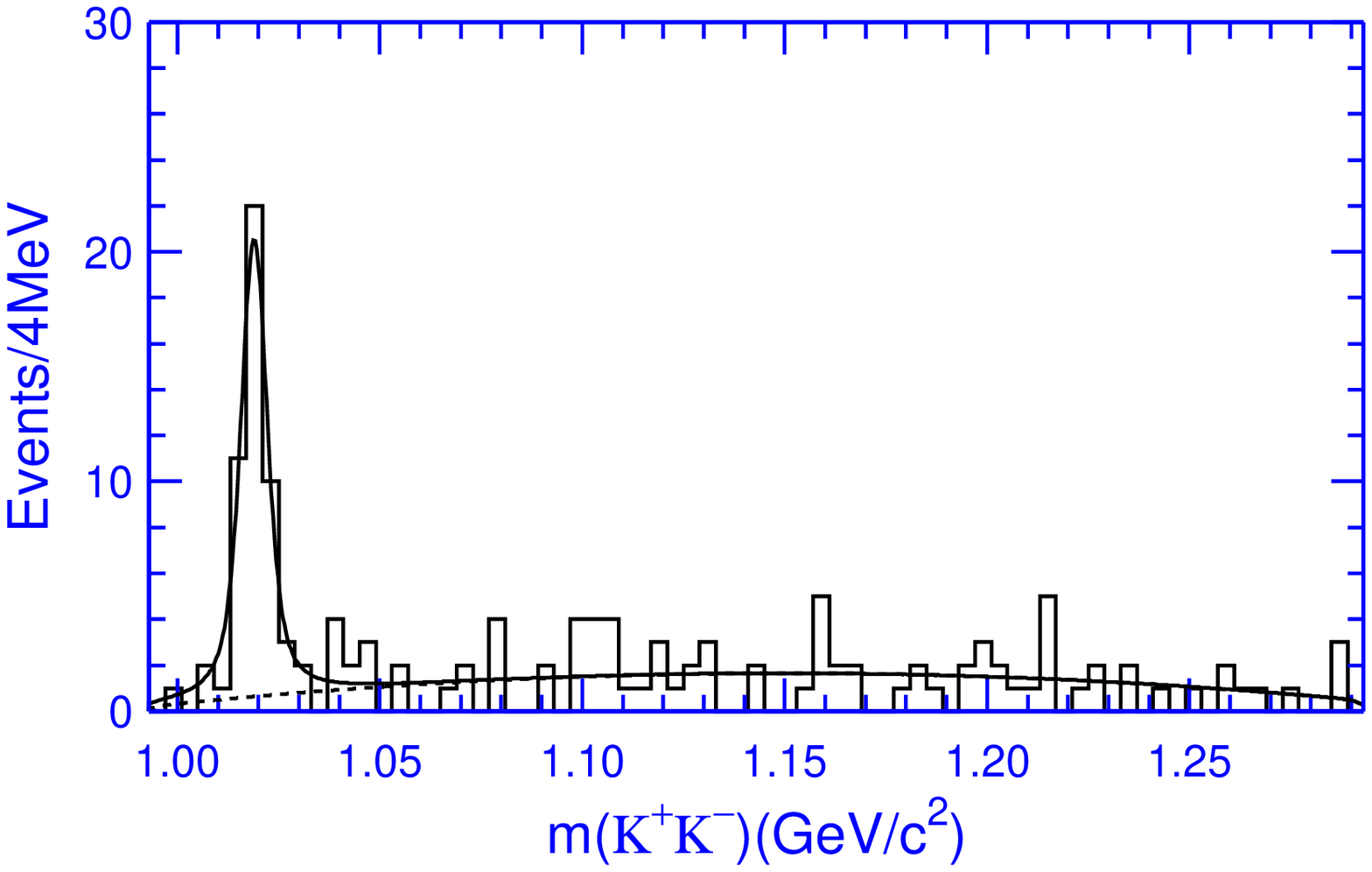,height=4.5cm,width=0.5\textwidth}}
\caption{\label{fig:phipipifit}
	The  $\kk$ invariant mass distribution for candidate
	$\psipto\fipi$ events. }
\end{figure}

 The candidate events for this decay mode have a final state $\kk\pipi$.
The combined probability for the assignment of $\psipto\kk\pipi$
is required to be larger than those of $\ppb\pipi$, $\pppp$, $\kkkk$, 
and $\kpipipi$.
A cut of $|\mrecpipi-m_{J/\psi}|>0.05$ GeV  rejects possible backgrounds 
from $\psipto\pipi\jpsi$. The decay of $\psipto K^*K^-\pi^+ 
(+c.c.)\rar\kk\pipi$ has a smooth $m_{\kk}$ distribution below 1.06 
GeV and therefore does not affect the $\fipi$ signal.  No $K_s^0$ signal is 
found in the $m_{\pipi}$ invariant mass distribution for the selected data 
sample, indicating negligible $K_s^0K^{\pm}\pi^{\mp}$ background.   
Figure \ref{fig:phipipifit}  shows the  $\kk$ invariant mass
distribution for $\psipto\fipi$ candidates, where a prominent $\phi$ signal 
can be seen. A fit gives $51.5\pm8.3$ signal events with the width 
of the $\phi$ fixed to its PDG value. The detection efficiency for this
decay mode is $17.8\%$, and we determine the branching fraction
$$B(\psipto\fipi)=(1.5\pm0.2\pm0.2)\times 10^{-4}.$$

\subsubsection{$\psipto \fifo$}
\begin{figure}[htbp]
\centerline{\epsfig{figure=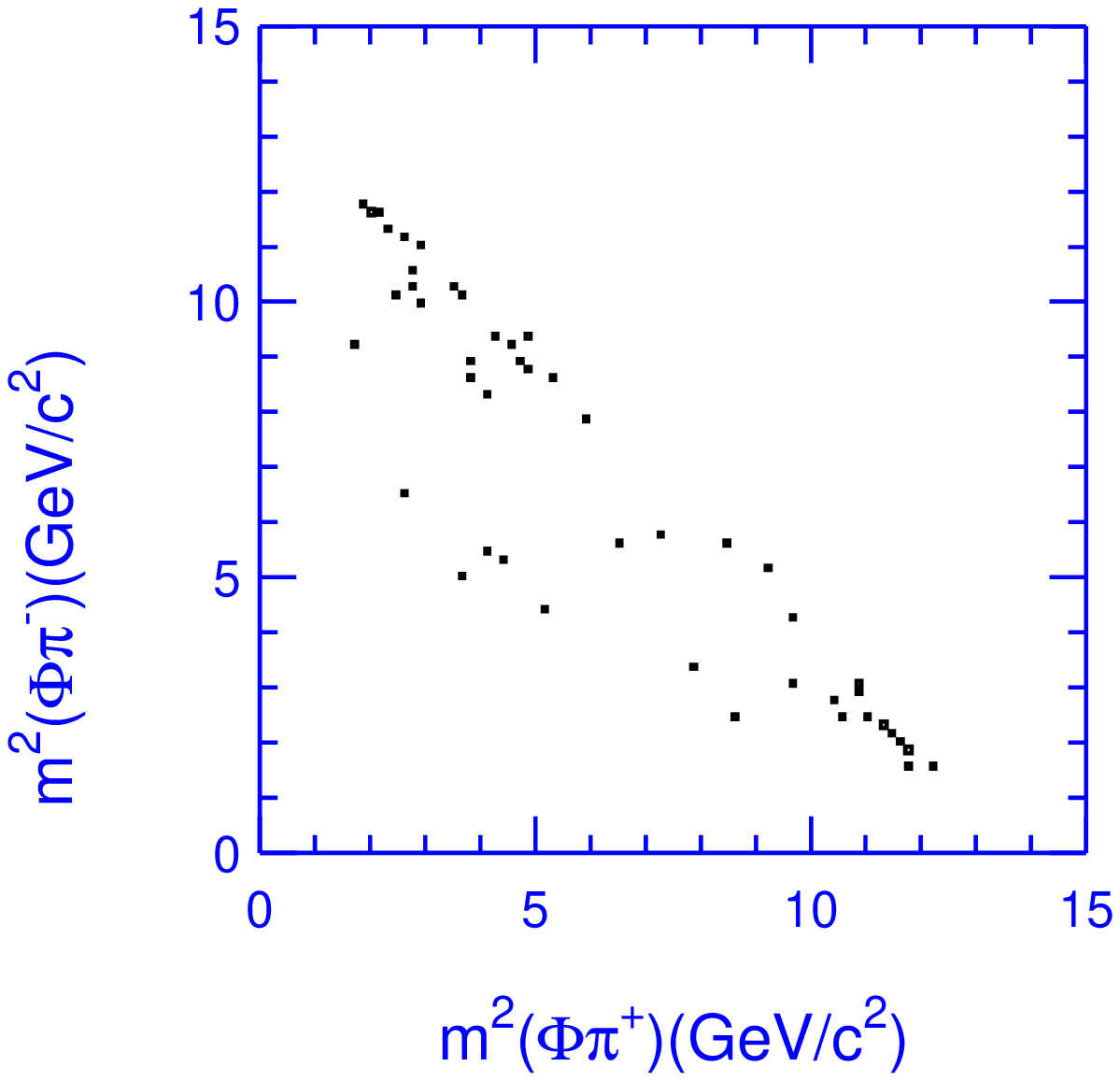,height=5cm,width=5cm}}
\caption{\label{fig:phif02d}
        The dalitz plot for candidate
        $\fipi$ events. }
\end{figure}

\begin{figure}[htbp]
\centerline{\epsfig{figure=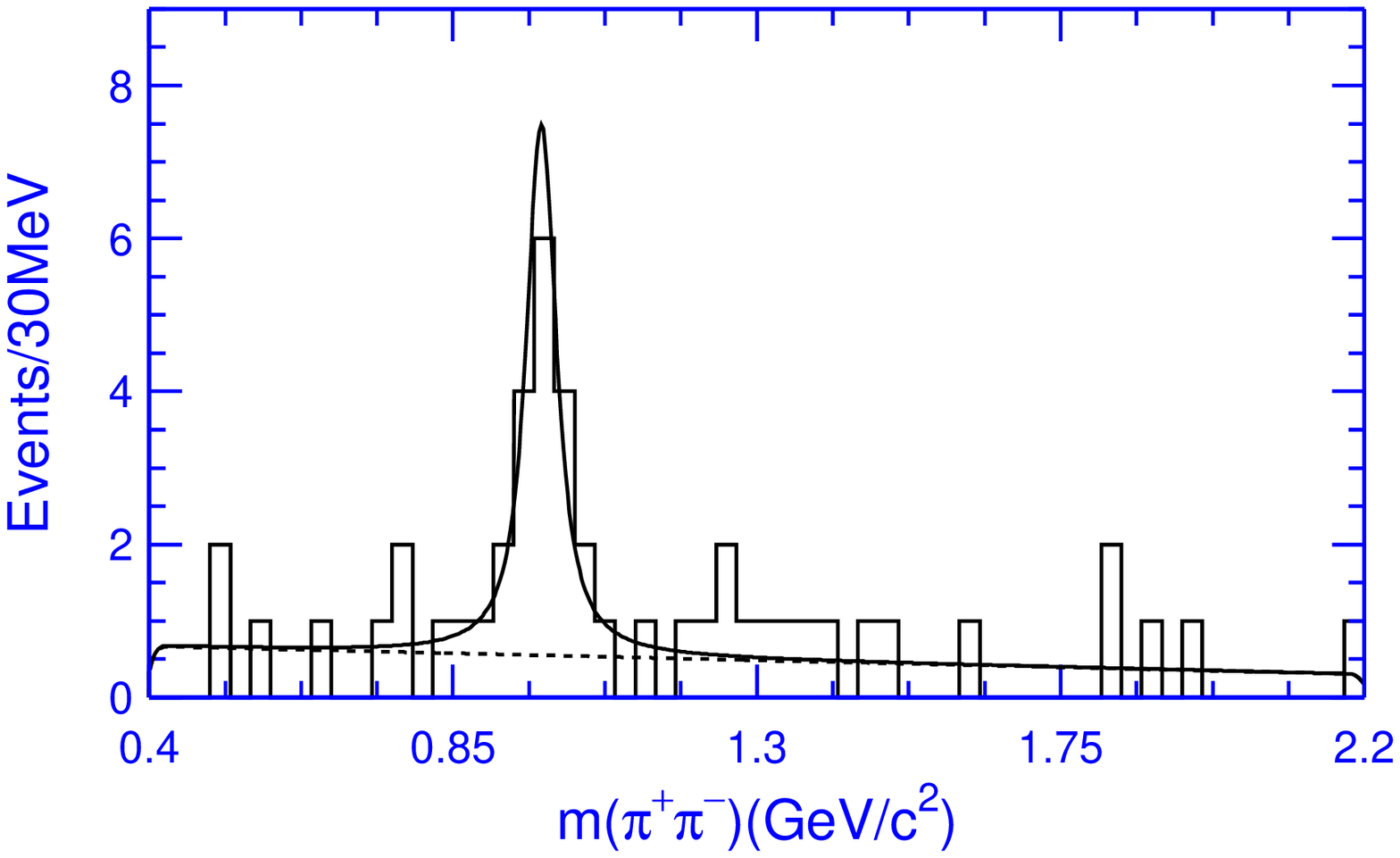,height=4cm,width=0.5\textwidth}}
\caption{\label{fig:phif0fit}
	The invariant mass distribution for candidate
	$\psipto\phi f_0(980)$ events. }
\end{figure}

    We use the same criteria as those for $\fipi$ for this 
decay mode, but with an additional requirement $|m_{\kk}-m_{\phi}|<0.02$ GeV to 
select events containing a $\phi$ particle. The dalitz plot is shown in 
Fig.~\ref{fig:phif02d}, which indicates the existence of the $\phi f_0(980)$
signal. Leaving the $f_0(980)$ width to float, the fit
to the $\pi^+\pi^-$ invariant mass shown in  Fig. \ref{fig:phif0fit}
 gives $18.4\pm6.4$ signal events
with the fitted $f_0(980)$ width of about 45 MeV,
the statistical significance is about $5.8\sigma$.
 The detection efficiency for this
decay mode is $17.0\%$, and we determine the branching fraction
$$B(\psipto\fifo)\cdot B(\fifo\rar\pipi)$$
$$=(0.6\pm0.2\pm0.1)\times 10^{-4}.$$

\subsubsection{$\psipto \fikk$}
\begin{figure}[htbp]
\centerline{\epsfig{figure=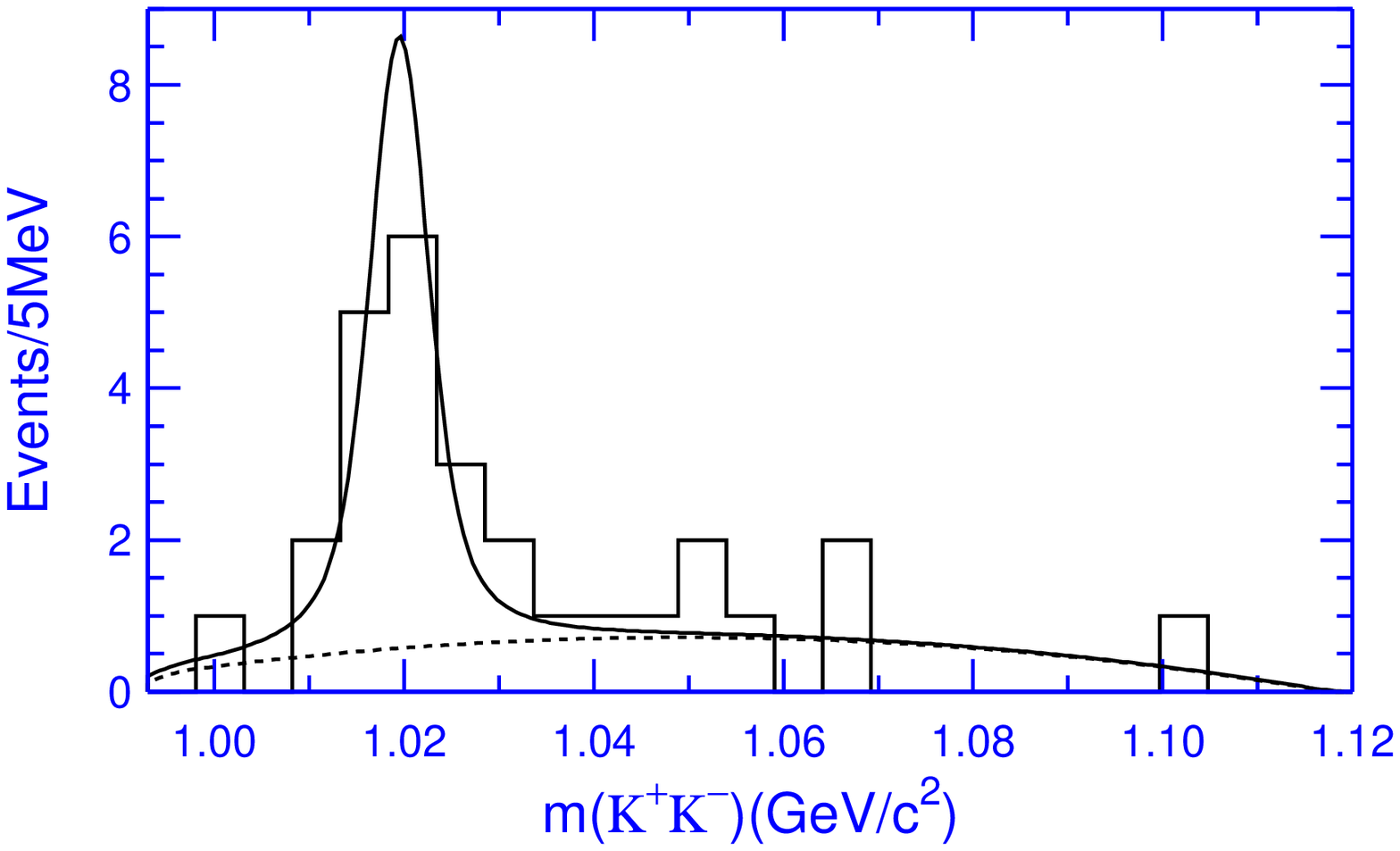,height=4.5cm,width=0.5\textwidth}}
\caption{\label{fig:phikkfit}
	The  $\kk$ invariant mass distribution for candidate
	$\psipto\fikk$ events. }
\end{figure}

Here the combined probability for the assignment of $\psipto\kkkk$
is required to be larger than those of $\ppb\kk$, $\kk\pipi$, and $\pppp$.
The $\kk$ invariant mass
distribution in Fig. \ref{fig:phikkfit} shows a clear 
$\phi$ peak. A fit gives $16.1\pm5.0$ signal events with the width
of the $\phi$ fixed to its PDG value. The detection efficiency for this
decay mode is $13.4\%$, and we determine the branching fraction
$$B(\psipto\fikk)=(0.6\pm0.2\pm0.1)\times10^{-4}.$$

\subsubsection{$\psipto \fipp$} 
\begin{figure}[htbp]
\centerline{\epsfig{figure=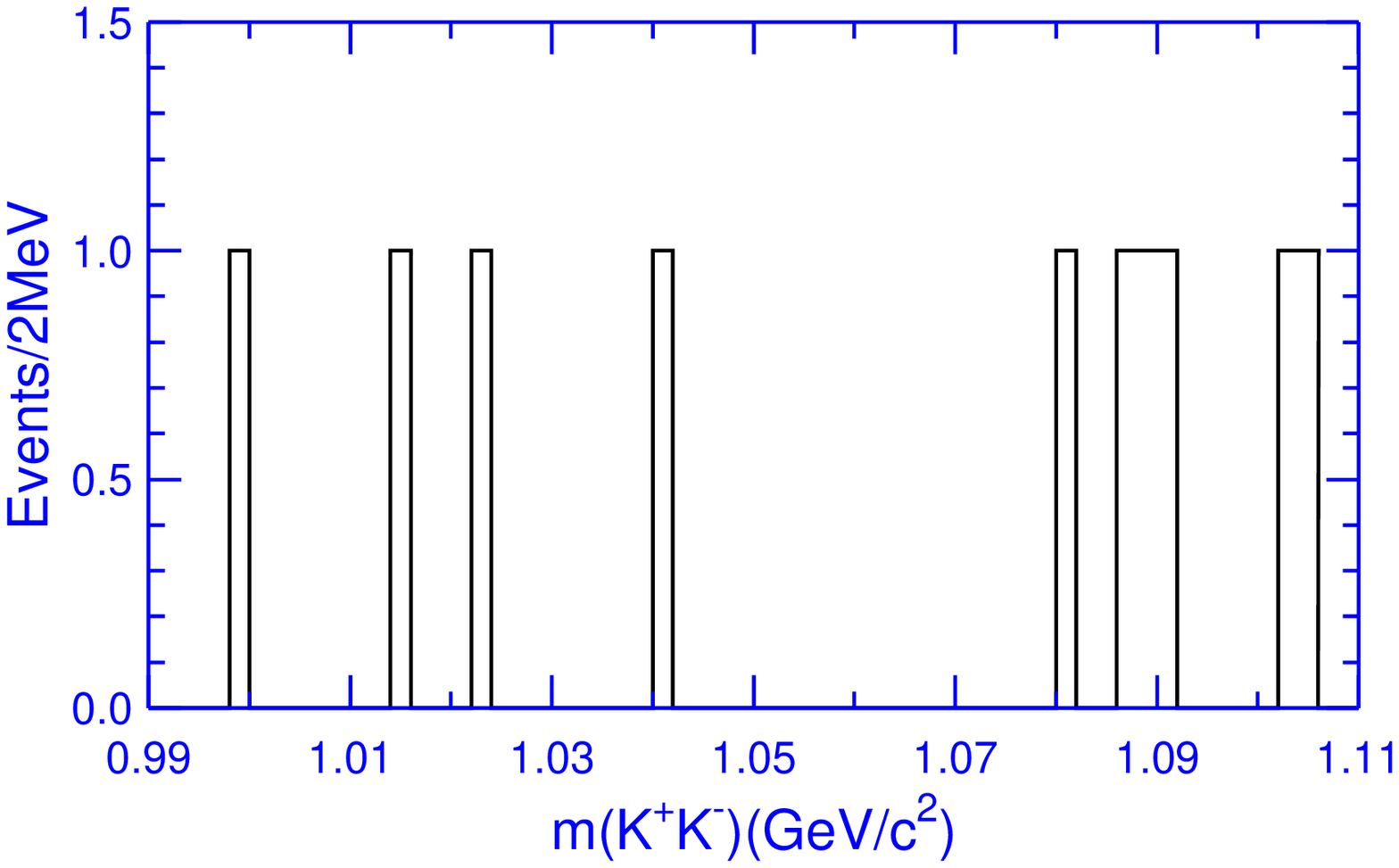,height=3.5cm,width=0.5\textwidth}}
\caption{\label{fig:phippfit}
	The  $\kk$ invariant mass distribution for candidate
	$\psipto\fipp$ events. }
\end{figure}

   The combined probability for the assignment of $\psipto\kk\ppb$
is required to be larger than those of $\ppb\pipi$, $\kkkk$, and $\pppp$.
The $\kk$ invariant 
mass plot is shown in Fig. \ref{fig:phippfit}. Only four events appear 
in the $\phi$ mass region.  Assuming zero background events and using a
detection efficiency of  $16.8\%$, 
 we obtain the upper limit on the branching fraction of 
$$B(\psipto\fipp)<0.26\times 10^{-4} ~~~~( 90\%~~ C.L.)$$

\section{Branching Fraction determination} \label{BFD}
  For a process $\psipto X$, the  branching fraction is 
determined by the relation
$$B(\psipto X)=$$
$$\frac{n^{obs}(\psipto X\rar Y)}
  {N_{\psip}\cdot B(X\rar Y)\cdot\epsilon(\psipto X\rar Y)},$$
where Y stands for the final state, X the intermediate state, 
and $\epsilon$ the detection efficiency. The branching fraction of 
$X\rar Y$ is taken from the PDG \cite{PDG}.  The total 
number of $\psip$ events 
$N_{\psip}=(4.02\pm0.22)\times 10^6$ \cite{Npsip} is determined from the 
number of 
$\psipto \pipi\jpsi$ events corrected for detection efficiency in the BES 
$\psip$ data sample ($1.227\pm0.003\pm0.017\times 10^6$) \cite{psip_no} 
and the PDG branching fraction \cite{PDG}.  

\subsection{Efficiency Corrections and Systematic errors} \label{sys}
Because the Monte Carlo does not simulate real events exactly, it is
necessary to correct the detection efficiency
obtained from simulation for the difference between MC and real data 
caused in PID and kinematic fitting. To correct for
the PID difference, the efficiency is multiplied by a factor ranging
from 0.89 to 0.98 with an uncertainty of 0.04 to 0.07, depending on
channel; while the correction factor in kinematic fitting is
$0.85\pm0.05$ and $0.85\pm0.08$ for 4-prong and 4-prong plus 2-photon
final states, respectively~\cite{Eff_Corr}.

Beside the uncertainties caused by the particle identification and the
kinematic fitting stated above, a systematic error common to all decay modes is the uncertainty in
the total number of $\psip$ events ($5.4\%$). The uncertainties of the
PDG values of the intermediate state $\omega$, $\phi$, $b_1$,
$f_2(1270)$, and $f_0(980)$ decay branching fractions are also sources
of systematic error ($0.8\%$ to $3.1\%$). 
%
The systematic error due to the statistical
precision of the MC event samples ranges from $1.2\%$ to $3.2\%$,
depending on the decay channel. 
Difficulties in the simulation of low energy
photons in the Monte Carlo give rise to a systematic error in the efficiency
that varies from $4.5\%$ to $8.6\%$ depending on photon energy for the final
states containing $\pi^0$.
The systematic error from
$\pi^0\rar 2\gamma$, where at least one photon is converted to a
$e^+e^-$ pair is about $1.4\%$. The variation of branching fraction
results for different choices of the fiducial region is about
$5\%$.  The total systematic error is taken as the sum of the
individual terms added in quadrature and ranges from $12\%$ to $17\%$,
depending on the channel.

\subsection{Branching fraction results}  \label{BF}  
   The results, including numbers of 
signal events, detection efficiencies and branching fractions or upper 
limits ($90\%$ C.L.), are summarized in Table~\ref{results}. The first error of the branching fraction is 
statistical and the second is systematic for each channel.  Among these, 
the branching fractions for $\omega f_{2}(1270)$ and $b_{1}^{\pm}\pi^{\mp}$ 
supersede previous BES results ~\cite{vt,bes_ap}. 
For  $b_{1}^{\pm}\pi^{\mp}$, the
difference is due to an improved understanding of the acceptance; for
$\omega f_{2}(1270)$, the difference is due to improved selection criteria to reduce
background. 

To test the $12\%$ rule, we also list in Table~\ref{results} the ratio 
$Q_h$ of the $\psip$ and $\jpsi$ 
branching fractions for each channel, where 
the $\jpsi$ branching fractions are taken from the PDG. Among these channels, 
the ratio of $\omega f_{2}(1270)$ (VT mode) is suppressed by a
factor of five with respect to the PQCD expectation, and
those of $\omega \pi^{+}\pi^{-}$, $\omega p\bar{p}$, and $\phi K^{+}K^{-}$ 
are suppressed by about a factor of two, 
those of other channels  
are consistent with PQCD expectation within errors
except $\phi p\bar{p}$ channel.

In conclusion, we have presented first measurements
of branching fractions for $\psip$ decays to  
 $\opi$,  $\ok$,$\opp$,  $\fipi$,$\phi f_0(980)$ , $\fikk$ and $\fipp$
channels, and supersede previous $b_1\pi$ and $\omega f_2(1270)$
results. This work further confirmed previous BES finding
that the suppression puzzle of the hadronic decays of the $\psip$ with
respect to the $J/\psi$ extends from the VP decay to VT decay, and the 
$b_1\pi$ (AP mode) decay is consistent with PQCD expectation. As to the VS
($\phi f_0(980)$) decay and $Vh\bar{h}$ three-body decays
(but $\fipp$,which needs more statistics), we have provided the first test
for the "12\% rule", which seems generally valid, although there might
be some fluctuations.

\acknowledgments
We acknowledge the strong efforts of the BEPC staff and the 
helpful assistance from the members of the IHEP 
computing center. The work of the BES Collaboration is supported in part by
the National Natural Science Foundation of China
under Contract No. 19991480, 10175060 and the Chinese Academy of Sciences
under contract No. KJ95T-03, 
and by the Department of
Energy under Contract Nos.
 DE-FG03-92ER40701 (Caltech),
DE-FG03-93ER40788 (Colorado State University), DE-AC03-76SF00515 (SLAC),
 DE-FG03-94ER40833 (U Hawaii),
DE-FG03-95ER40925 (UT Dallas).

\begin{table*}
\caption{\label{results}
	 Branching fractions of $\psi(2S)$ and $Q_h$ values for $\psi(2S)$ and
	$J/\psi$ hadronic decays.$^{*1}$ }
\vspace{0.5cm}
\begin{ruledtabular}
\begin{tabular}{l c c c c c c } 
  Channel   &  Number of  & Efficiency & $\frac{B_{\psi(2S)\rightarrow h}}{B_{\psi(2S)\rightarrow \pi^+\pi^- J/\psi}}$ 
            & $B_{\psi(2S)\rightarrow h}$ & $B_{J/\psi\rightarrow h}$ & $Q_h (\%)$      \\ 
  ~~~~h     &    Events   &   ($\%$)   & ($10^{-4}$)  & ($10^{-4}$) &  $(10^{-4})$ & $(\%)$ \\ \hline
$\omega \pi^{+}\pi^{-}$   & $100\pm12$     & $5.8\pm0.8$
                          & $15.8\pm1.9\pm2.2$ & $4.8\pm0.6\pm0.7$     
                          & $72.0\pm12.0$  & $6.7\pm1.7$  \\ \hline 
$b_{1}^{\pm}\pi^{\mp}~^{*2}$    & $61\pm11$   & $5.2\pm0.7$
                          & $10.6\pm1.9\pm1.5$ & $3.2\pm0.6\pm0.5$ 
                          & $30.0\pm5.0$   & $11\pm3$ \\ \hline
$\omega f_{2}(1270)~^{*2}$      & $10.2\pm4.9$    & $4.8\pm0.7$
                          & $3.4\pm1.7\pm0.5$ & $1.1\pm0.5\pm0.2$       
                          & $43.0\pm6.0$   & $2.4\pm1.3$  \\
                                & & & & $<1.5$                \\ \hline
$\omega K^{+}K^{-}$       & $23.0\pm5.2$   & $4.4\pm0.6$
                          & $4.8\pm1.1\pm0.7$ & $1.5\pm0.3\pm0.2$           
                          & $7.4\pm2.4$    & $20\pm8$   \\ \hline
$\omega p\bar{p}$         & $14.9\pm5.8$   & $5.4\pm0.8$
                          & $2.5\pm1.0\pm0.4$ & $0.8\pm0.3\pm0.1$           
                          & $13.0\pm2.5$   & $6.0\pm2.8$     \\ \hline 
$\phi\pi^+\pi^-$              & $51.5 \pm8.3 $ & $17.8\pm2.1$
                          & $4.8\pm0.8\pm0.6$ & $1.5\pm0.2\pm0.2$             
                          & $8.0\pm1.2$    & $18\pm5$    \\ \hline
$\phi f_0(980)(f_0\rightarrow\pi^+\pi^-)^{*3}$  & $18.4\pm6.4 $    & $17.0\pm2.1 $
                          & $1.8\pm0.6\pm0.2$ & $0.6\pm0.2\pm0.1$              \\
$\phi f_0(980)^{*4}$      & & 
                          & $3.4\pm1.2\pm0.4$ & $1.1\pm0.4\pm0.1$ &  $3.2\pm0.9$
&$33\pm15$    \\ \hline
$\phi K^{+}K^{-}$         & $16.1\pm5.0 $  & $13.4\pm1.6 $
                          & $2.0\pm0.6\pm0.2$ & $0.6\pm0.2\pm0.1$        
                          & $8.3\pm1.3$    & $7.3\pm2.6$      \\ \hline
$\phi p \bar{p}$          & $4 $           & $16.8\pm1.8 $
                              &$<0.85$ & $<0.26$ &$0.45\pm0.15$ &$<58$        \\ 
\end{tabular}
\end{ruledtabular}
{*1} The upper limit  is at the $90\%$ confidence level; $B_{J/\psi}$ taken from
PDG value. \\ 
{*2} $b_{1}^{\pm}\pi^{\mp}$ and $\omega f_{2}(1270)$ events are subsets
of $\omega \pi^+\pi^-$ events. \\
{*3} $\phi f_0(980)$ events  are subset of $\phi\pi^+\pi^-$ events.\\
{*4} $B_{f_0\rightarrow\pi^+\pi^-}=0.521\pm0.016$(PDG'96) 
\end{table*}

\end{document}